\documentclass[aip,jap,letterpaper,amsmath,amssymb,reprint]{revtex4-1}
\usepackage{graphicx}
\usepackage{microtype}
\usepackage[utf8]{inputenc}

\begin{document}
\title{Giant perpendicular exchange bias with antiferromagnetic MnN}
\affiliation{Center for Spinelectronic Materials and Devices, Department of Physics, Bielefeld University, D-33501 Bielefeld, Germany}
\author{P. Zilske}
\email{pzilske@physik.uni-bielefeld.de}
\author{D. Graulich}
\author{M. Dunz}
\author{M. Meinert}

\date{\today}

\begin{abstract}
We investigated an out-of-plane exchange bias system that is based on the antiferromagnet MnN. Polycrystalline, highly textured film stacks of Ta / MnN / CoFeB / MgO / Ta were grown on SiO$_x$ by (reactive) magnetron sputtering and studied by x-ray diffraction and Kerr magnetometry. Nontrivial modifications of the exchange bias and the perpendicular magnetic anisotropy were observed both as functions of film thicknesses as well as field cooling temperatures. In optimized film stacks, a giant perpendicular exchange bias of 3600\,Oe and a coercive field of 350\,Oe were observed at room temperature. The effective interfacial exchange energy is estimated to be $J_\mathrm{eff} = 0.24$\,mJ/m$^2$ and the effective uniaxial anisotropy constant of the antiferromagnet is $K_\mathrm{eff} = 24$\,kJ/m$^3$. The maximum effective perpendicular anisotropy field of the CoFeB layer is $H_\mathrm{ani} = 3400$\,Oe. These values are larger than any previously reported values. These results possibly open a route to magnetically stable, exchange biased perpendicularly magnetized spin valves.
\end{abstract}

\maketitle

Spin electronics allows to realize nonvolatile fast low-power computer memory and is well established in hard disk drive read heads and magnetic sensors.\citep{Gregg2002, Chappert2007} The key component in spin electronic devices, a magnetoresistive element using either giant magnetoresistance (GMR) or tunnel magnetoresistance (TMR), is composed of two magnetic films: a free sense layer and a fixed reference layer. The magnetization of the ferromagnetic free layer follows external magnetic fields or can be switched by a current via the spin transfer torque. The reference layer has to be stable against external fields to allow for different magnetic alignments of the two layers, which give rise to the magnetoresistance. The reference layer is typically created by pinning a thin ferromagnetic (FM) film to an antiferromagnetic (AFM) film via the exchange bias (EB) effect.\citep{Meiklejohn1957, Schuller1999, Berkowitz1999, Stamps2001, Kiwi2001, Ohldag2003, Grady2010} In a typical device, the magnetic hysteresis loop of the reference layer is shifted by the exchange bias to fields that are not encountered during normal device operation.

Thin films with perpendicular magnetic anisotropy (PMA) are of great interest for spintronic devices. The tunable anisotropy energy allows to enhance the thermal stability of the magnetization and lower critical current densities for the spin-transfer torque switching are achievable as compared to in-plane magnetized systems.\citep{Nishimura2002, Meng2006, Mangin2006} Thus, interest in systems showing perpendicular EB (PEB) increased as well. There are several studies about (Co/Pt)$_n$ and (Co/Pd)$_n$ multilayer systems coupled with an AFM such as IrMn or FeMn.\citep{Maat2001, Garcia2003, Sort2005, Ji2006, Liu2009} However, the reported perpendicular exchange bias field values $H_\mathrm{eb}$ are similar to the coercive field $H_\mathrm{c}$, making these systems not attractive for practical applications that require $H_\mathrm{eb} \gg H_\mathrm{c}$. Chen \textit{et al.} investigated CoFe/IrMn bilayers with ultrathin CoFe yielding PEB values up to 1000\,Oe at room temperature (RT).\citep{Chen2014} Moreover, Zhang \textit{et al.} observed large PEB up to 1145\,Oe in CoFeB/MgO systems pinned by MnIr,\citep{Zhang2015} which are the highest reported values in PEB systems up to now.

In the present article, we report on an exchange bias system that is based on antiferromagnetic MnN. It crystallizes in the $\theta$-phase of the Mn-N phase diagram,\citep{Gokcen1990} which crystallizes in the tetragonal variant of the NaCl structure with $a = 4.256$\,\AA{} and $c = 4.189$\,\AA{} at room temperature where the precise numbers depend on the N content in the material. With increasing nitrogen content, larger lattice constants are observed. \citep{Otsuka1977, Suzuki2000, Leineweber2000} The N\'eel temperature of MnN is about 660\,K; \citep{Leineweber2000} the magnetic transition is accompanied by a tetragonal-to-cubic structural transformation.\citep{Gokcen1990} The magnetic order of the material was investigated with neutron powder diffraction and by first-principles calculations.\citep{Leineweber2000, Suzuki2001, Lambrecht2003} It was found to be collinear of AFM-I type, i.e., with the magnetic moments coupled parallel within the $ab$ planes and alternating along the $c$ direction. The spin orientation is under debate; Leineweber et al. found it to be along the $c$ direction just below the N\'eel temperature, whereas it would tilt slightly away from the $c$ axis at a lower temperature.\citep{Leineweber2000} Instead, Suzuki et al. found the spin direction to be in the $ab$ planes.\citep{Suzuki2001} An important difference between these experiments lies in the preparation procedures, that led to a slightly nitrogen-poor $\theta$-MnN in the first case, whereas the $\theta$-MnN was saturated in the second case, which is also reflected by the larger lattice constants in the latter case. We therefore propose that the data measured by Suzuki et al. reflect the intrinsic properties of the stoichiometric $\theta$-MnN phase. In both cases, the Mn magnetic moments were found to be 3.3 $\mu$B at room temperature.

In a recent article, we reported on in-plane exchange bias in Ta / MnN / CoFe / TaO$_x$ grown on thermally oxidized Si substrates.\cite{Meinert2015} Large exchange bias up to 1800\,Oe at room temperature was observed with an effective interfacial exchange energy of $J_\mathrm{eff} = t_\mathrm{CoFe} M_\mathrm{CoFe} \mu_0 H_\mathrm{eb} = 0.41$\,mJ/m$^2$, an effective uniaxial anisotropy constant of $K_\mathrm{eff} = J_\mathrm{eff} / t_\mathrm{crit} = 37$\,kJ/m$^3$, and a median blocking temperature of 160$^\circ$C. Here, we go beyond our previous report by using ultrathin CoFeB as the ferromagnet to obtain a perpendicular magnetization and investigate the perpendicular exchange bias. We prepared film stacks of Ta 10 nm / MnN $t_\mathrm{MnN}$ / Co$_{40}$Fe$_{40}$B$_{20}$ $t_\mathrm{CoFeB}$ / MgO 2 nm / Ta 0.5 nm / Ta$_2$O$_5$ 2 nm on thermally oxidized Si wafers by (reactive) dc magnetron sputtering at room temperature. We followed exactly the same preparation procedure as in our previous report. The base pressure of the system was around $5 \times 10^{-9}$\,mbar prior to the deposition runs. The MnN films were sputtered from an elemental Mn target in a mixed Ar and N$_2$ atmosphere with various working pressures and partial pressure ratios. Optimization of the deposition parameters with respect to the exchange bias yielded a 50:50 N$_2$:Ar mixture at \hbox{$p = 2.3 \times 10^{-3}$\, mbar} as the best deposition condition. The typical deposition rate of MnN was 0.1 nm/s at a source power of 50 W. Subsequent post-annealing and field cooling in a magnetic field of $H_\mathrm{FC} = 6.5$\, kOe perpendicular to the film plane was performed in a vacuum furnace. Magnetic characterization of the stacks was performed using the polar and longitudinal magneto-optical Kerr effect (MOKE) at room temperature. Structural and film thickness analyses were performed with a Philips X’Pert Pro MPD, which is equipped with a Cu source and Bragg-Brentano optics. 

\begin{figure}
\centering
\includegraphics[width=8cm]{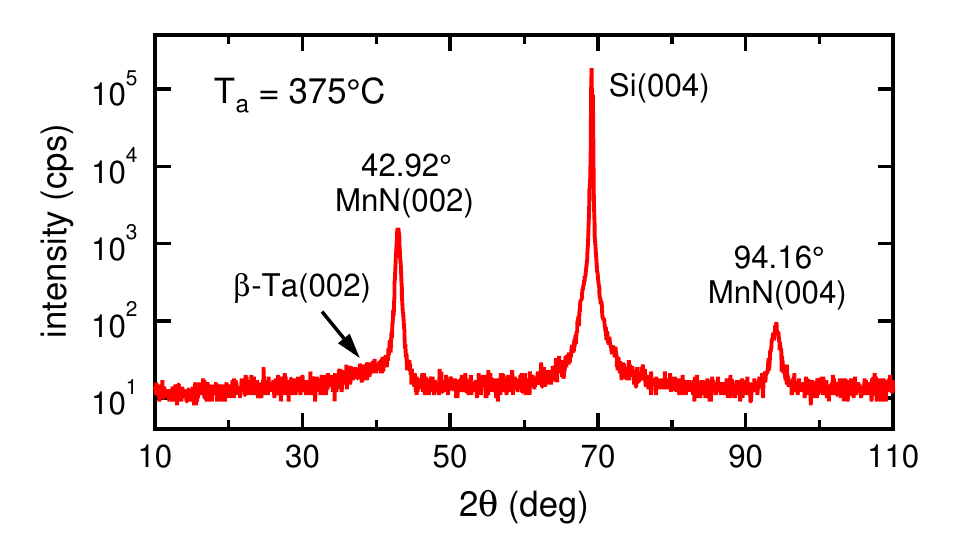}
\caption{X-ray diffraction spectrum of a Ta 10\,nm/MnN 30\,nm/CoFeB 0.65\,nm/MgO 2\,nm/Ta 0.5\,nm/Ta$_2$O$_5$ 2\,nm film annealed at 375$^\circ$C for 15 minutes.}
\label{mnn_xrd}
\end{figure}

To verify the growth of the $\theta$-phase of MnN we performed x-ray diffraction analysis of MnN films with varying film thicknesses and annealing conditions. In Figure \ref{mnn_xrd} a typical x-ray diffraction spectrum of the Ta / MnN / CoFeB / MgO stacks is shown. All peaks are identified as belonging to the substrate, to the $\beta$-Ta seed layer, or to the MnN film. The lattice constant of MnN after annealing is 4.21 \AA{}, which is slightly larger than the bulk values reported in the literature. No other phases are observed, and the lattice constant is too large to identify the structure as the cubic $\epsilon$-Mn$_4$N or the tetragonal $\eta$-M$_3$N$_2$ phase \citep{Yang2002}. The measurements revealed a polycrystalline, columnar growth of the MnN in (001) direction. A detailed structural characterization of the MnN is reported in our previous article.\citep{Meinert2015}

\begin{figure}
\centering
\includegraphics[width=8.5cm]{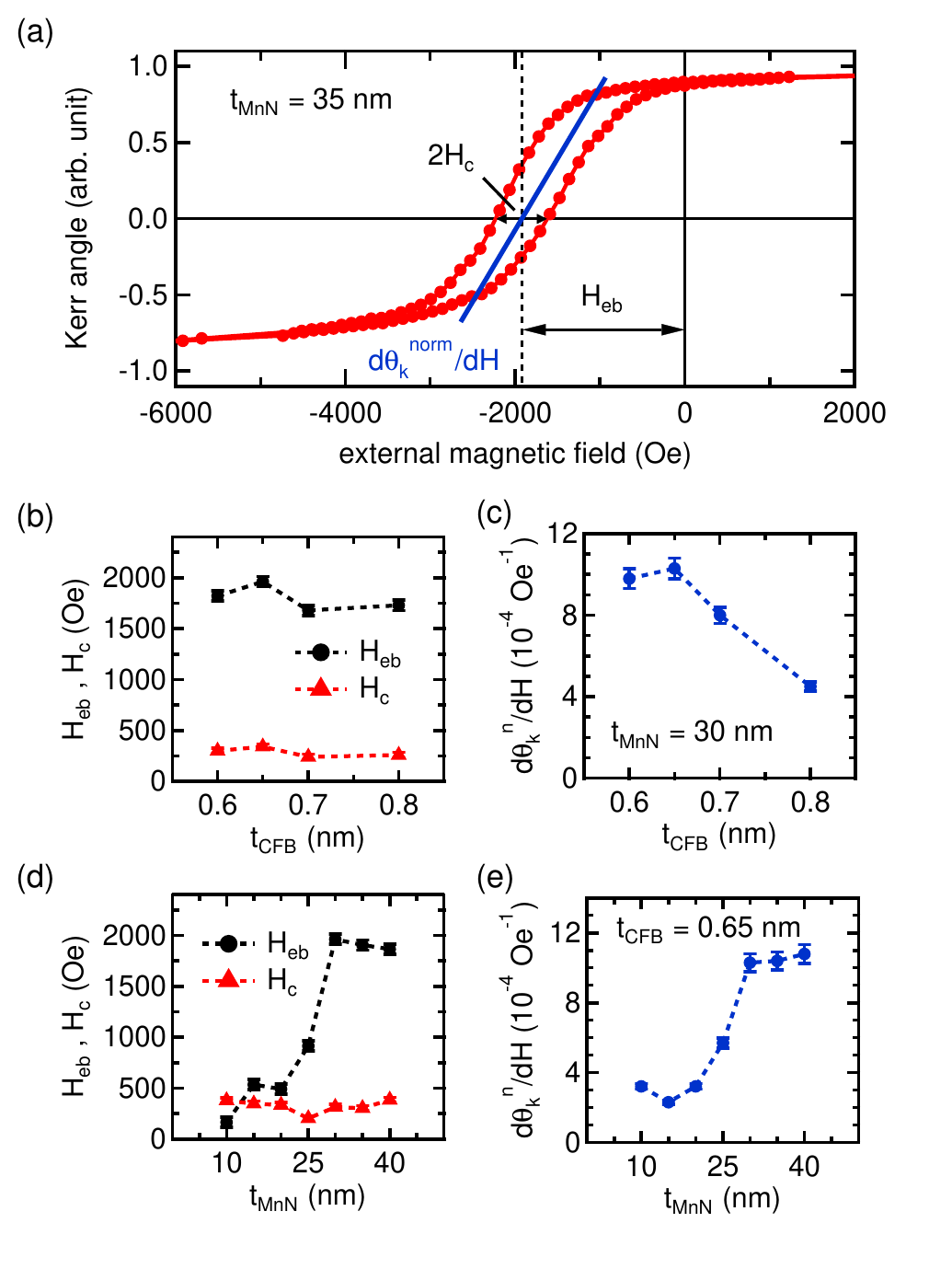}
\caption{(a) Normalized polar MOKE loop detected with the magnetic field perpendicular to the sample plane for a film with $t_\mathrm{MnN} = 35$\,nm and $t_\mathrm{CoFeB} = 0.65$\,nm. (b) Dependence of $H_\mathrm{eb}$ and $H_\mathrm{c}$ on the CoFeB thickness for $t_\mathrm{MnN} = 30$\,nm. (c) Dependence of d$\theta_\mathrm{k}^\mathrm{norm}/$d$H \rvert_{H=H_\mathrm{eb}}$ on the CoFeB thickness for $t_\mathrm{MnN} = 30$\,nm. (d) Dependence of $H_\mathrm{eb}$ and $H_\mathrm{c}$ on the MnN thickness for $t_\mathrm{CoFeB} = 0.65$\,nm. (e) The dependence of d$\theta_\mathrm{k}^\mathrm{norm}/$d$H \rvert_{H=H_\mathrm{eb}}$ on the MnN thickness for $t_\mathrm{CoFeB} = 0.65$\,nm. The samples were annealed at 375$^\circ$C for 15 minutes. Dashed lines are guide to the eye.}
\label{mnn_fast}
\end{figure}

In Figure \ref{mnn_fast}(a) we show a perpendicular magnetic hysteresis loop of an optimized MnN / CoFeB / MgO stack with $t_\mathrm{MnN} = 35$\,nm and $t_\mathrm{CoFeB} = 0.65$\,nm. The definitions of the exchange bias field $H_\mathrm{eb}$ and the coercive field $H_\mathrm{c}$ are indicated by arrows. Additionally, the derivative of the normalized Kerr rotation $\mathrm{d}\theta_\mathrm{k}^\mathrm{norm}/\mathrm{d}H$ at $H=H_\mathrm{eb}$ is indicated in the figure. We use this quantity as a simple and quick estimate of the PMA, which we later demonstrate to be valid. The loop shows high exchange bias and a reasonably low coercive field with $\mathrm{d}\theta_\mathrm{k}^\mathrm{norm}/\mathrm{d}H \rvert_{H=H_\mathrm{eb}} \approx 10^{-3}\,\mathrm{Oe}^{-1}$. The FM is almost saturated at zero external field. In the following we discuss the variation of the film thicknesses and post-annealing conditions for the MnN/CoFeB/MgO stacks to maximize the exchange bias field as well as the derivative d$\theta_\mathrm{k}^\mathrm{norm}/$d$H \rvert_{H=H_\mathrm{eb}}$ and simultaneously minimize the coercive field of the CoFeB. All denoted exchange bias values are determined under consideration of the training effect, which only occurs for small MnN thicknesses $t_\text{MnN} < 15$\,nm.

\begin{figure}
\centering
\includegraphics[width=8.4cm]{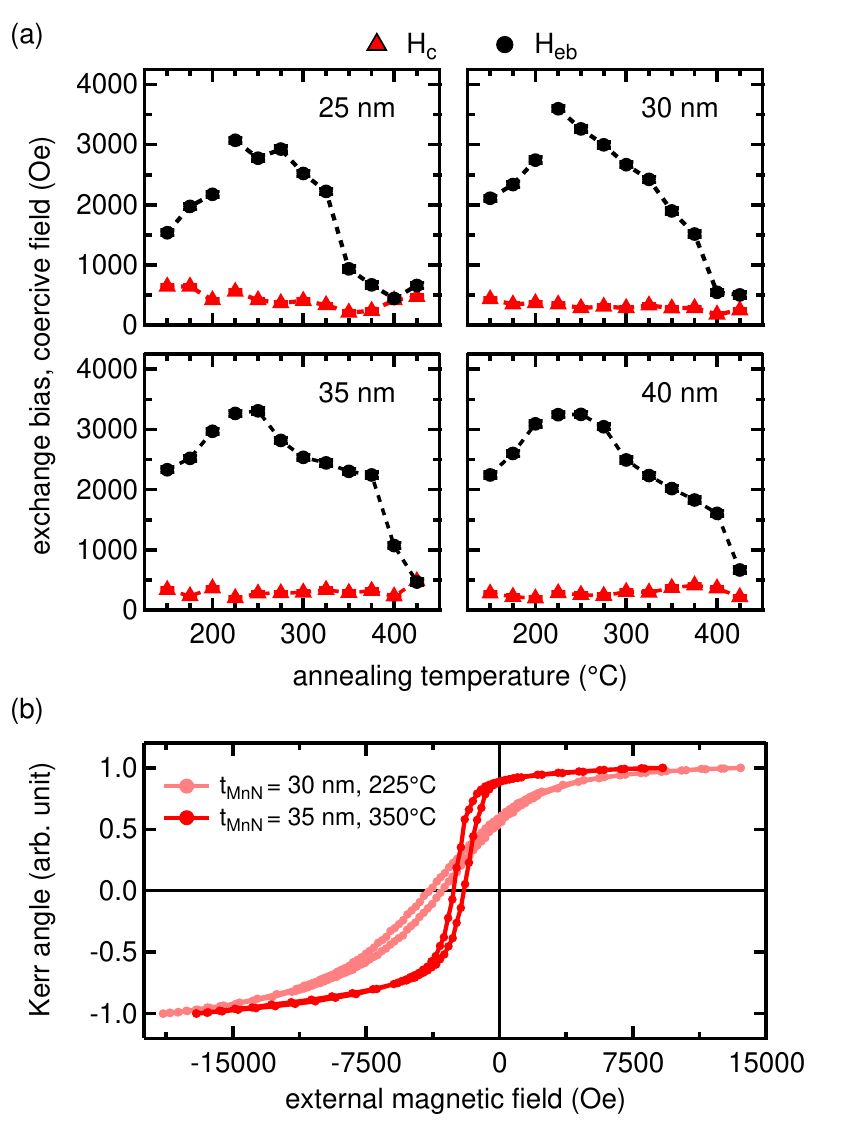}
\caption{(a) Dependence of $H_\mathrm{eb}$ and $H_\mathrm{c}$ on the annealing temperature for MnN thicknesses of $t_\mathrm{MnN}=25,30,35,40$\,nm. The CoFeB thickness was $t_\mathrm{CoFeB}=0.65$\,nm. The samples were successively annealed for 60 minutes. (b) Corresponding polar MOKE loops detected with the magnetic field perpendicular to the sample plane for a film with $t_\mathrm{MnN} = 30$\,nm annealed at 225$^\circ$C and a film with $t_\mathrm{MnN} = 35$\,nm annealed at 350$^\circ$C.}
\label{temp_2540}
\end{figure}

\begin{figure}
\centering
\includegraphics[width=8.5cm]{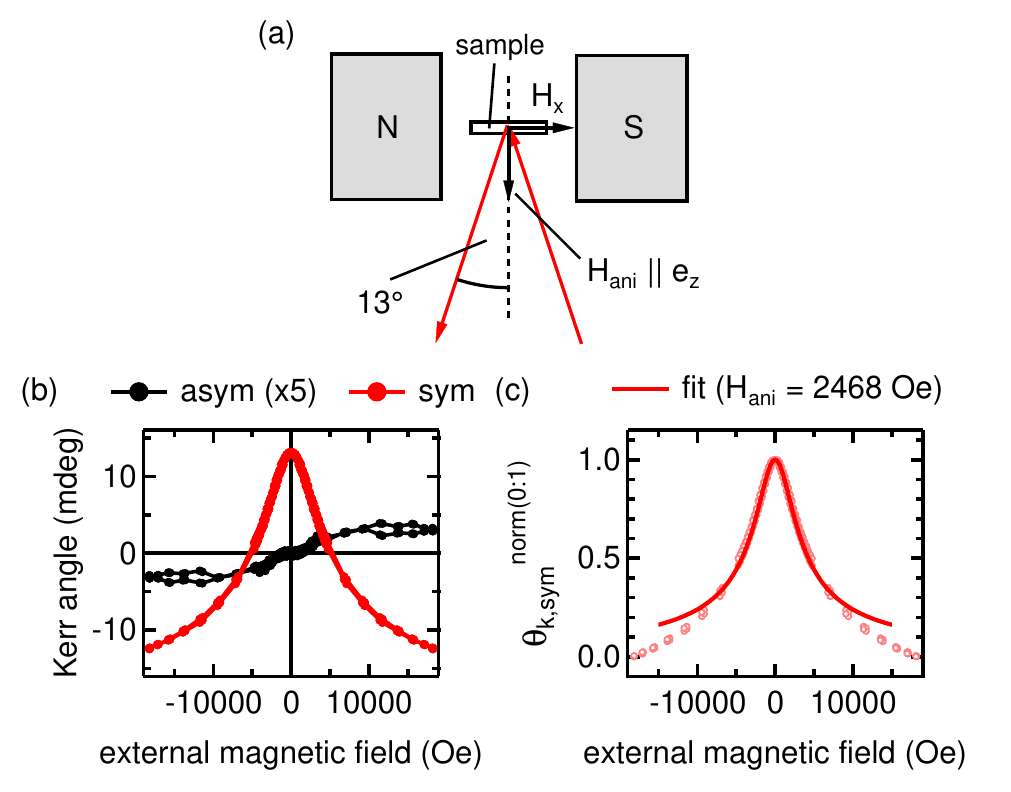}
\caption{(a) Longitudinal MOKE setup geometry with the magnetic field in the sample plane. The anisotropy field is assumed to point out of the sample plane. (b) Symmetric and antisymmetric component of the detected MOKE data. Due to the difference in amplitude, the antisymmetric data are multiplied by a factor of five. (c) Normalized, symmetric MOKE component with the corresponding fit function yielding an anisotropy field of $H_\text{ani}=(2468\pm 21)$\,Oe. The film thickness is $t_\text{MnN}=30$\,nm and the annealing temperature $T_\text{A}=300^\circ$C.}
\label{ip_ausw}
\end{figure}

Figure \ref{mnn_fast}(b) displays the variation of the exchange bias field $H_\mathrm{eb}$ and the coercive field $H_\mathrm{c}$ with the CoFeB thickness $t_\mathrm{CoFeB}$. For the CoFeB thickness variation a weak maximum of $H_\mathrm{eb}$ was observed at $t_\mathrm{CoFeB} = 0.65$\, nm. The derivative $\mathrm{d}\theta_\mathrm{k}^\mathrm{norm}/\mathrm{d}H \rvert_{H=H_\mathrm{eb}}$ exhibits a sharp maximum for this CoFeB thickness, as can be seen in Figure \ref{mnn_fast}(c). Because of that, $t_\mathrm{CoFeB} = 0.65$\,nm was fixed for the following investigations. The variation of $H_\mathrm{eb}$ and $H_\mathrm{c}$ with $t_\mathrm{MnN}$ is shown in Figure \ref{mnn_fast}(d). For all thicknesses $t_\mathrm{MnN} \geq 15$\,nm the exchange bias field is larger than the coercive field which is important for the potential integration into magnetic devices. The coercive field $H_\mathrm{c}$ lies between 250 and 400\,Oe following no specific behavior. The critical thickness of the MnN $t_\mathrm{crit}$ for the onset of exchange bias is about 10\,nm and the maximum exchange bias is found at $t_\mathrm{MnN}=30$\,nm. Thus, the MnN thickness dependence of $H_\mathrm{eb}$ resembles very closely the previously reported in-plane case with the same critical thickness and the same thickness for the maximum exchange bias.  A very similar film thickness dependence can be observed for the derivative $\mathrm{d}\theta_\mathrm{k}^\mathrm{norm}/\mathrm{d}H \rvert_{H=H_\mathrm{eb}}$, cf. Figure \ref{mnn_fast}(e). This gives direct evidence for the intimate link between exchange bias and magnetocrystalline anisotropy and demonstrates that the anisotropy of the antiferromagnet provides a significant contribution to the total PMA of the stack.\cite{Grady2010} In the following we will focus on the samples with large $H_\mathrm{eb}$ and large PMA, i.e., $t_\mathrm{MnN} \geq 25$\,nm.

In Figure \ref{temp_2540}(a) we show the annealing temperature dependence of the exchange bias and the coercive field for MnN thicknesses $t_\mathrm{MnN}=25, 30, 35, 40$\,nm. In all cases, the coercive field does not change significantly with the annealing temperature. In contrast, the exchange bias shows a strong dependence on the the annealing temperature. For all thicknesses, the maximum exchange bias is observed for annealing temperatures around 225$^\circ$C, yielding giant values up to 3600\,Oe for $t_\mathrm{MnN}=30$\,nm. Exchange bias of more than 3000\,Oe can be obtained for all thicknesses. For higher annealing temperatures the exchange bias decreases rapidly for thicknesses of $t_\mathrm{MnN}=25, 30$\,nm. For larger film thicknesses, this decrease at high annealing temperature is less pronounced and apparently the maximum temperature which still provides exchange bias is increased at larger film thickness. This is probably related to nitrogen diffusion into the Ta buffer layer, which degrades the MnN at high annealing temperature. Two corresponding hysteresis loops are shown in Figure \ref{temp_2540}(b).

Taking the previous results together, we can estimate the maximum effective interfacial exchange coupling energy $J_\mathrm{eff} = t_\mathrm{CoFeB} M_\mathrm{s} \mu_0 H_\mathrm{eb}$ and the effective uniaxial anisotropy constant $K_\mathrm{eff} = J_\mathrm{eff} / t_\mathrm{crit}$. We consider the typical saturation magnetization of CoFeB as $M_\mathrm{s} = 1000\,$kA/m. At $t_\mathrm{CoFeB} = 0.65$\,nm we obtain a maximum exchange bias field of $H_\mathrm{eb} = 3600$\,Oe, so the maximum interfacial exchange energy is $J_\mathrm{eff} = 0.24$\,mJ/m$^2$. With the critical thickness $t_\mathrm{crit} = 10$\,nm the effective uniaxial anisotropy at room temperature is $K_\mathrm{eff} = 24$\,kJ/m$^3$. Both of these values are smaller than the corresponding values of the in-plane magnetized system of Ta / MnN / CoFe / Ta from our previous report.\cite{Meinert2015} In comparison to other previously reported out-of-plane systems, the interfacial exchange energy is at least twice as big as the measured values of Sort \textit{et al.} and Chen \textit{et al.} ($J_\mathrm{eff} \sim 0.1$\,mJ/m$^2$).\citep{Sort2005, Chen2014} Furthermore, the energy is as big as the highest reported values of Liu \textit{et al.} ($J_\mathrm{eff} = 0.25$\,mJ/m$^2$) and Zhang \textit{et al.} ($J_\mathrm{eff} = 0.21$\,mJ/m$^2$).\citep{Liu2009, Zhang2015} 

\begin{figure}
\centering
\includegraphics[width=8.8cm]{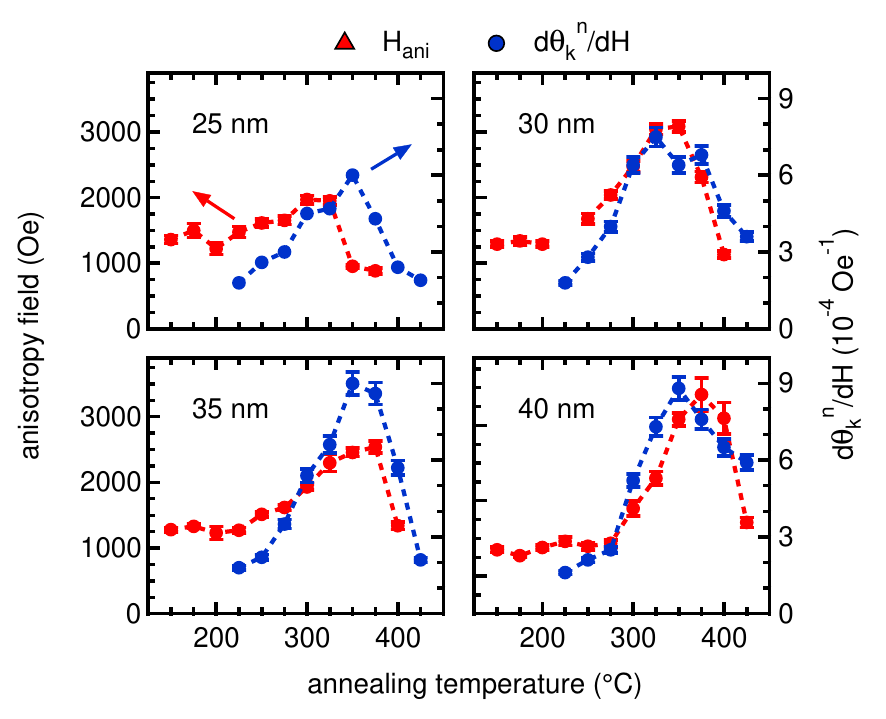}
\caption{Dependence of the anisotropy field (left axis) and the derivative d$\theta_\mathrm{k}^\mathrm{norm}/$d$H \rvert_{H=H_\mathrm{eb}}$ (right axis) on the annealing temperature for MnN thicknesses of $t_\mathrm{MnN}=25,30,35,40$\,nm. The CoFeB thickness was $t_\mathrm{CoFeB}=0.65$\,nm. The samples were successively annealed for 60 minutes.}
\label{ip_2540}
\end{figure}

Finally, the effective anisotropy fields were measured by longitudinal MOKE with the magnetic field in the sample plane. The symmetric component of the hysteresis loop is normalized and can be expressed for $H_\mathrm{ani} \gg H_\mathrm{demag}$ as 
\begin{align}
\theta_\text{sym}^\mathrm{norm[0:1]}(H_\mathrm{x}) &= \frac{H_\text{ani}}{\sqrt{H_\text{ani}^2+H_\mathrm{x}^2}}.
\label{one}
\end{align}
This way, the anisotropy field can be determined by fitting the normalized, symmetric component of the longitudinal MOKE data. The corresponding MOKE setup geometry, as well as exemplary MOKE data are shown in Figure \ref{ip_ausw}. The results of the analysis are depicted in Figure \ref{ip_2540} together with the values of $\mathrm{d}\theta_\mathrm{k}^\mathrm{norm}/\mathrm{d}H \rvert_{H=H_\mathrm{eb}}$. For all thicknesses we observe the same trend: for annealing temperatures up to around 250$^\circ$C, the anisotropy field is around 1400\,Oe. After annealing at temperatures between 325$^\circ$C and 375$^\circ$C, the anisotropy field reaches its maximum value independent of the MnN thickness. The absolute maximum is about 3400\,Oe for $t_\mathrm{MnN}=40$\,nm at 375$^\circ$C. For higher temperatures, the anisotropy field decreases rapidly for all thicknesses. In comparison, the anisotropy field for $t_\mathrm{MnN}=25$\,nm is much smaller than for the other thicknesses, and also does not show a strong variation with the annealing temperature. As already mentioned, d$\theta_\mathrm{k}^\mathrm{norm}/$d$H \rvert_{H=H_\mathrm{eb}}$ is directly proportional to the anisotropy field for most thicknesses and annealing temperatures.

In conclusion, we prepared Ta / MnN / CoFeB / MgO stacks with giant perpendicular exchange bias up to 3600\,Oe. The optimized thicknesses are $t_\text{CoFeB}=0.65$\,nm and $t_\text{MnN} \geq 30$\,nm. Combining the annealing temperature dependence of the exchange bias and the anisotropy field, we can sum up that in a range between 325$^\circ$C and 375$^\circ$C high anisotropy fields around 3000\,Oe as well as high exchange bias around 2500\,Oe are achieved. This makes MnN a promising candidate for integration into perpendicular magnetic tunnel junctions or perpendicular spin valves.

We acknowledge financial support by the Ministerium für Innovation, Wissenschaft und Forschung des Landes Nordrhein-Westfalen (MIWF NRW). We further thank G. Reiss for making available laboratory equipment.

\end{document}